# Small FDIRC Designs

B. Dey[+], B. Ratcliff,[*] and J. Va'vra[*]

+   Sezione INFN di Milano, Milano, Italy
*   SLAC National Accelerator Laboratory, CA 94025, USA

*Abstract* – In this article[1] we explore the angular resolution limits attainable in small FDIRC designs taking advantage of the new highly pixelated detectors that are now available. Since the basic FDIRC design concept attains its particle separation performance mostly in the angular domain as measured by two-dimensional pixels, this paper relies primarily on a pixel-based analysis, with additional chromatic corrections using the time domain, requiring single photon timing resolution at a level of 100-200ps only. This approach differs from other modern DIRC design concepts such as TOP or TORCH detectors,[2] whose separation performances rely more strongly on time-dependent analyses. We find excellent single photon resolution with a geometry where individual bars are coupled to a single plate, which is coupled in turn to a cylindrical lens focusing camera.

### INTRODUCTION

We have previously built and tested a full scale prototype of a Focusing DIRC (FDIRC) detector [1]. This device was based on the BaBar DIRC radiators [2] attached to a new cylindrically focused camera, and was intended for the upgrade of the BaBar detector for the SuperB factory. Similar optical concepts are now being considered for the GLUEX experiment at JLAB [3], and possibly, the Electron-Ion collider PID detector [4]. Several BaBar bar boxes remain in storage and we hope that this work may encourage consideration of using them in other applications.

The SuperB FDIRC detector was studied in detail [1] at the SLAC cosmic ray telescope using 14 Hamamatsu H-8500 MaPMTs with 6mm x 6mm pixels. The telescope provided 3D tracking, which allowed successful testing of the concept and the electronics. However, due to the wide range of momenta for the cosmic muons and somewhat limited angular tracking resolution (~1.5 mrad), the FDIRC prototype achieved a single photon Cherenkov resolution of ~10 mrad, which is about 2 mrad worse than what could have achieved in a test beam.

In this paper, we probe the Cherenkov angular resolutions attainable in three different designs using FDIRC style cylindrical lens focusing as follows:

I.   a modified SuperB FDIRC camera design using 3 mm x 12 mm pixels (H-9500) and unmodified BaBar bar boxes;
II.  a new smaller FDIRC focusing block (FBLOCK) than that used in the original SuperB design (and design I) with 3 mm x 12 mm pixels (Hamamatsu H-9500) and unmodified BaBar bar boxes;
III. the same as (II), but with a modified BaBar bar box which replaces the last of the 4(x12) short bars that make up the twelve 4.9 meter long bars filling each box with a one meter-long plate spanning the whole bar box.[3] The motivation is to greatly reduce the effect of the bar width on the pin-hole Cherenkov angle resolution in the x-direction, where there is no lens (mirror) focusing.

The paper will compare and contrast the single photon angular performance attained by each scheme, using GEANT4 based Monte Carlo simulations. In the simulations, we assume 10 GeV muon tracks, impinging perpendicularly on the bar surface. Single photon timing resolution is assumed to be less than 100 ps. This simulation was tested earlier on the SuperB FDIRC design [1], and found to be consistent with experimental data.

### SuperB FDIRC design (I) with small pixels

The SuperB optics is shown on Fig.1a. It uses an unmodified BaBar bar box, and couples to a new wedge and FBLOCK, which was designed for detectors with 6 mm$^2$ pixels (for example the H-8500 MaPMT). In this paper, we consider smaller pixel sizes of 3 mm x 12 mm pixel using the Hamamatsu H-9500 MaPMT (four small pixels are connected together in x-direction).[4]

**(a)**

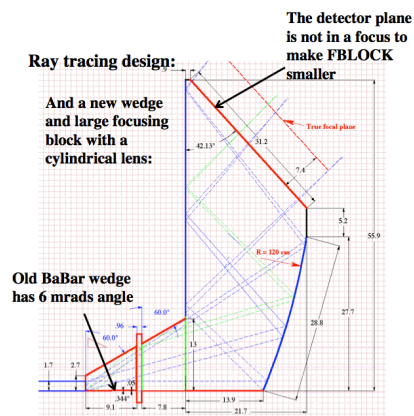

---

This work supported by the Department of Energy, contract DEAC02-76SF00515.

[1] The paper was presented at the RICH 2016 conference, Bled, Slovenia, September 2016.

[2] See presentation at this conference.

[3] A single Babar bar box has a set of 12 narrow 4.88 m-long Fused Silica bars, glued out of four shorter segments, each 1.22 m-long. In this simulation for design with the plate, we replace one group of short bars, closest to FBLOCK, with a 1 m long plate, filling the entire width of the bar box.

[4] One should note that in this SuperB design, the detector was not in the focal plane in order to reduce the size of the FBLOCK.

**Small FDIRC design (II) with smaller FBLOCK**

Fig.2a shows an alternative FDIRC design with a smaller FBLOCK volume (about half the volume of the SuperB FBLOCK design). One should note that in this new design the detector plane is in focus. **This design uses the BaBar bar boxes without any modification.** The FBLOCK is coupled to the BaBar box via a new small wedge, similar to the original SuperB design. Detectors are assumed to be Hamamatsu H-9500 MaPMTs with 3 mm x 12 mm pixel sizes (again four small pixels are connected together in the x-direction). The detector plane is smaller, thus reducing the number of PMT's required for full coverage, which also reduces costs.

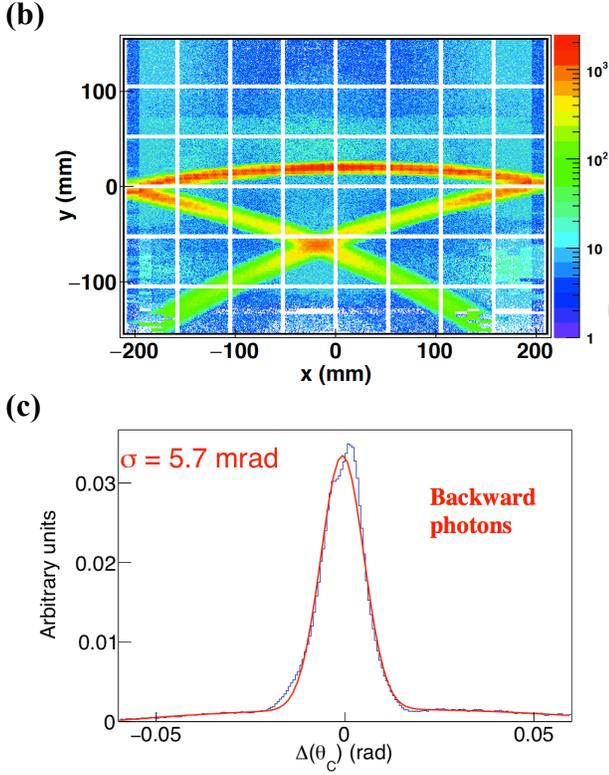

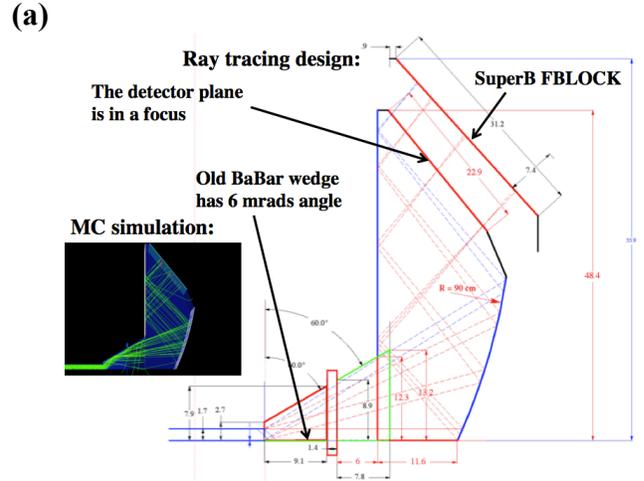

**Fig.1** (a) SuperB FDIRC design with cylindrical mirror focusing and 3 mm x12 mm pixels, as provided by the 256-pixel H-9500 MaPMT, with 4 pixels connected in the x-direction. (b) Cherenkov x-y hit pattern for perpendicular 10GeV muon tracks.[5] (c) Cherenkov single photon angle resolution for the entire Cherenkov ring and backward-going photons (10 GeV perpendicular muon tracks with |dTOP| < 2ns).[6] The black histogram shows the MC data, as fit by the red curve.

Fig.1b shows the Cherenkov ring x-y hit pattern for perpendicular tracks. One can see that the best resolution is in the central region of the ring (~4.2 mrad), and it gets progressively worse towards edges due to the kaleidoscopic effect [5]. Fig.1c shows the single photon Cherenkov angle resolution, after the timing-based chromatic correction, for backward-going photons. In this paper we define forward-going photons as photons going toward FBLOCK, and backward-going photons as going to the bar-end and reflecting back towards the FBLOCK.

In this MC analysis we use the variable dTOP = "$TOP_{MC\_time}$ - $TOP_{calculated}$", where $TOP_{MC\_time}$ is the time from the MC gun till the pixel hit, and $TOP_{calculated}$ is the calculated photon time-of-propagation from the pixel look-up table.[7] After the chromatic correction, we make a loose cut of |dTOP| < 2ns. Examples of dTOP distributions can be found in [1].

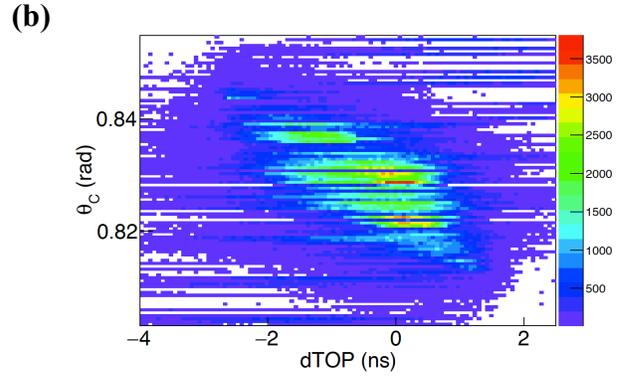

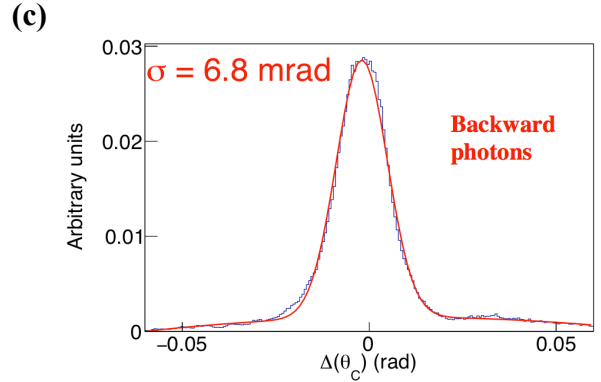

**Fig.2** (a) Smaller FDIRC design (II) with a reduced FBLOCK size (about 2x smaller volume), with 3 mm x 12 mm pixel sizes, as provided by the 256-pixel H-9500 MaPMT. (b) Correlation due to the chromatic effect between the Cherenkov angle and dTOP time for backward going photons. (c) Single photon Cherenkov angle resolution over the entire ring for backward going photons after the chromatic correction (10 GeV perpendicular muon tracks with |dTOP| < 2ns). The black histogram shows the MC data, as fit by the red curve..

---

[5] The reason for the asymmetry is that the MC gun was placed above the center of bar 6, i.e., not quite symmetrically along the width of the bar box.

[6] The forward-going photons propagate to the FBLOCK directly; the backward-going photons propagate to the bar box end, reflect from a mirror and then travel back to the FBLOCK.

[7] We have also tried a variable dTOP/Lpath, where Lpath is a total photon pathlength inside the bar box. No significant improvement was found.

Fig.2b shows the correlation due to the chromatic effect between the Cherenkov angle and dTOP for backward-going photons. Fig.2c shows the single photon Cherenkov angle resolution for backward photons, without the chromatic correction. The resolution is comparable to the Fig.1c resolution.

We have also evaluated a design based on the Photonis XP-85022 MCP-PMT, arranged to provide 1.6 mm x 25.6 mm size pixels; this did not yield any significant improvement in the resolution.

## Small FDIRC design (III) with a plate in a modified bar box

Fig.3a shows a third FDIRC design (III), where the bar boxes have been modified as described above (see Footnote 3). The first two FDIRC designs (I and II) used unmodified Babar bar boxes. However, since the FBLOCK is focusing only in the y-direction, and the total propagation length in the camera is rather short, the bar width makes a significant contribution to the Cherenkov angle resolution in the pin hole imaging system being utilized. The plate design, on the other hand, makes the x-contribution of individual bars to the Cherenkov angle resolution much smaller, since the pin-hole standoff distance is so long. Since the bar box must be modified, a new wedge is added which replaces the two wedges required in designs I and II. The FBLOCK is the same as that used in design II. The pixel sizes remain 3 mm x 12 mm as before.

Fig.3b shows the x-y hit pattern, which is more complicated than Fig.1b. One can see that the ring is folded more in the x-direction than in the earlier designs and compressed in the y-direction. We note that the detector plane itself has shrunk in height, but the y-axis range remains the same as in Fig.1b, for consistency. Figs.3c and 3d show the correlation due to the chromatic effect between the Cherenkov angle and dTOP time for backward going photons, before and after the correction. Comparing Fig.3c with Fig.2b, one can see clearly a benefit of the plate-based design, where the correlation is tighter. Figs.3e and 3f show the single photon Cherenkov angle resolution over the entire ring for backward and forward going photons respectively after the chromatic correction. The resolution is indeed excellent.

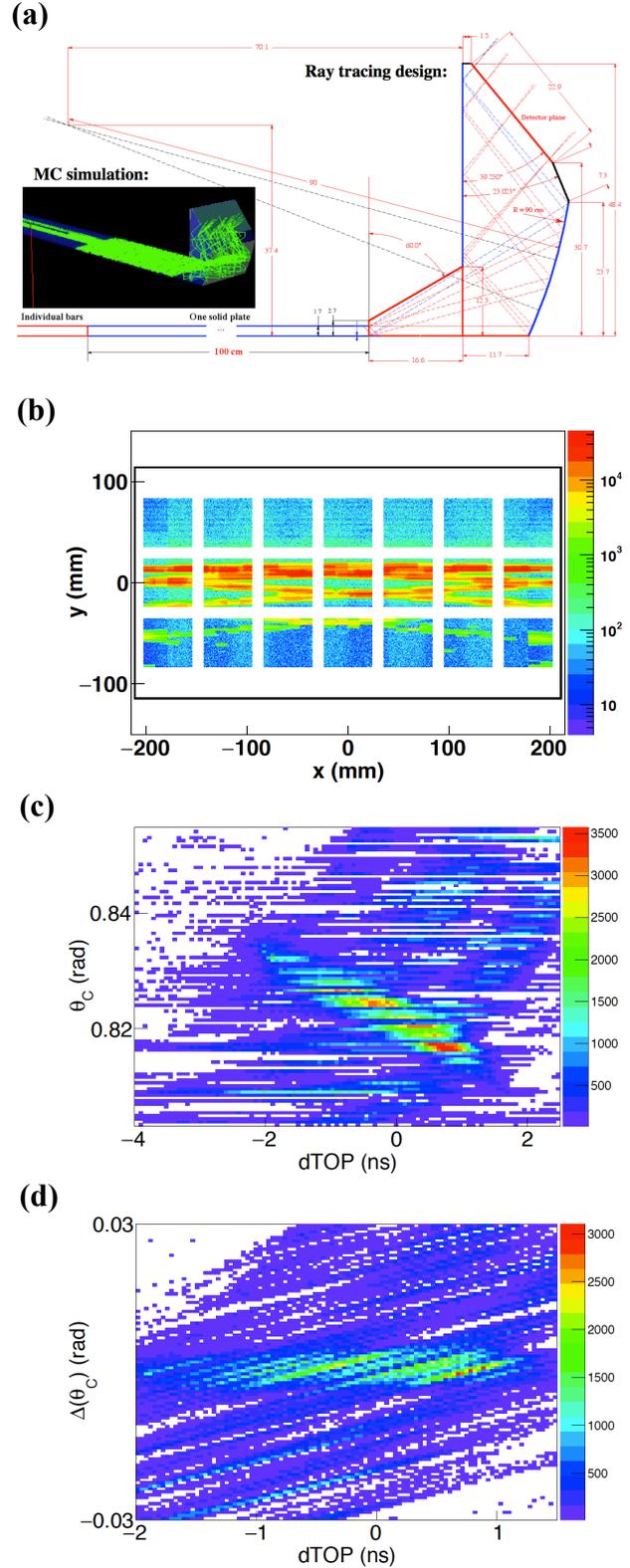

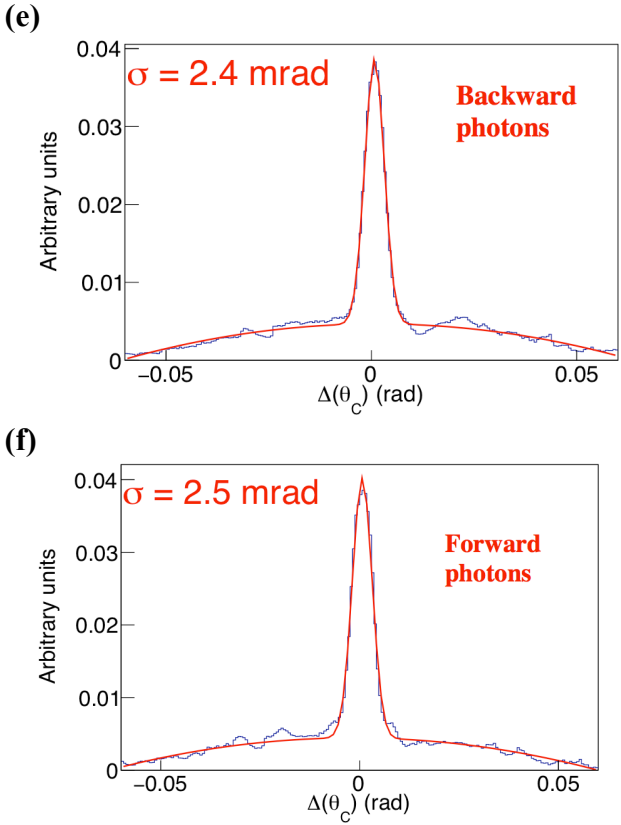

**Fig.3** (a) FDIRC design III with modified bar box, where the last group of twelve 1.22 meter-bars that make up the twelve 4.9 meter-long bars is replaced by a single meter-long 42 cm wide plate. The old BaBar bar box wedge and the additional one required in designs I and II are replaced by a single longer wedge. (b) Cherenkov ring as seen in this kind of design. (c&d) Correlation due to the chromatic effect between the Cherenkov angle and dTOP time for backward going photons, (c) before and (d) after the correction. Single photon Cherenkov angle resolution, after the chromatic correction was applied, for (e) backward and (f) forward going photons (10 GeV perpendicular muon tracks). The black histogram shows the MC data, as fit by the red curve.

## CONCLUSION

Building on our previous R&D on the focusing DIRC concept [1], we have explored the possibility of further enhancing the single photon resolution of such a device. We have considered several variants of both the focusing optics and the radiator bars with the following summarized results:

I. a SuperB-like FDIRC design with 3 mm x 12 mm pixels would provide a very good single photon Cherenkov angle resolution of ~6 mrad,[8]
II. a modified SuperB-like FDIRC design with a smaller FBLOCK with smaller 3 mm x 12 mm pixels would perform equally well, and be more cost effective since it requires less quartz material and a smaller number of MAPMTs,
III. a new FDIRC design with a combination of old DIRC bars coupled to a plate, and with 3 mm x 12 mm pixels, would provide the best single photon Cherenkov angle resolution of the DIRC concepts studied. However, it requires substantial effort to remake the bar boxes.

We hope this work will help to motivate the employment of mirror-based focusing with pixel-based DIRC devices in near future experiments, and perhaps suggest new uses for existing components such as the DIRC Bar Boxes. DIRC detectors that combine direct angular measurement with full use of the separation available in the time domain are likely to provide the "ultimate performance" available for this entire class of detectors.

---

[8] All resolution results in this paper are based on MC simulations. In realistic detectors there may be additional effects, which were not included.